\documentclass[nofootinbib,twocolumn,showpacs,preprintnumbers,pre,aps]{revtex4-1}

\usepackage[dvipdfmx]{graphicx}
\usepackage{bm}
\usepackage{amsmath}
\usepackage{amssymb}
\usepackage{txfonts}
\usepackage{color}
\usepackage{tikz}
\usepackage{esint}

\begin{document}

\title{Time-correlation functions of stochastic three-sphere micromachines}

\author{Jun Li}
\affiliation{Department of Physics, Wenzhou University, Wenzhou, Zhejiang 325035, China}
\affiliation{Wenzhou Institute, University of Chinese Academy of Sciences, 
Wenzhou, Zhejiang 325001, China}

\author{Ziluo Zhang}
\affiliation{Wenzhou Institute, University of Chinese Academy of Sciences, 
Wenzhou, Zhejiang 325001, China}
\affiliation{Institute of Theoretical Physics, Chinese Academy of Sciences, 
Beijing 100190, China}

\author{Zhanglin Hou}
\affiliation{Wenzhou Institute, University of Chinese Academy of Sciences, 
Wenzhou, Zhejiang 325001, China} 
\affiliation{Oujiang Laboratory, Wenzhou, Zhejiang 325000, China}

\author{Kento Yasuda}
\affiliation{Research Institute for Mathematical Sciences, 
Kyoto University, Kyoto 606-8502, Japan}

\author{Linli He}\email{Corresponding author: linlihe@wzu.edu.cn}
\affiliation{Department of Physics, Wenzhou University, Wenzhou, Zhejiang 325035, China}

\author{Shigeyuki Komura}\email{Corresponding author: komura@wiucas.ac.cn}
\affiliation{Wenzhou Institute, University of Chinese Academy of Sciences, 
Wenzhou, Zhejiang 325001, China} 
\affiliation{Oujiang Laboratory, Wenzhou, Zhejiang 325000, China}
\affiliation{
Department of Chemistry, Graduate School of Science,
Tokyo Metropolitan University, Tokyo 192-0397, Japan}

\date{\today}

\begin{abstract}
We discuss and compare the statistical properties of two stochastic three-sphere micromachines, 
i.e., odd micromachine and thermal micromachine.
We calculate the steady state time-correlation functions for these micromachines and decompose 
them into the symmetric and antisymmetric parts.
In both models, the cross-correlation between the two spring extensions has an antisymmetric part, which 
is a direct consequence of the broken time-reversal symmetry. 
For the odd micromachine, the antisymmetric part of the correlation function is proportional to the odd elasticity,
whereas it is proportional to the temperature difference between the two edge spheres for the thermal
micromachine.
The entropy production rate and the Green-Kubo relations for the two micromachines are also obtained.
Comparing the results of the two models, we argue an effective odd elastic constant of the thermal 
micromachine.
We find that it is proportional to the temperature difference among the spheres, which causes an 
internal heat flow and leads to directional locomotion in the presence of hydrodynamic interactions.
\end{abstract}

\maketitle

\section{Introduction}
\label{Sec:Int}

Microswimmers are tiny objects moving in fluids, such as sperm cells or motile bacteria,
that swim in a fluid and are expected to be relevant to microfluidics and microsystems~\cite{Lauga09a,LaugaBook,Hosaka22}.
By transforming chemical energy into mechanical work, microswimmers change their shapes and can move in 
viscous environments.
According to Purcell's scallop theorem, reciprocal body motion cannot be used for locomotion in a Newtonian
fluid~\cite{Purcell1977,Ishimoto12}.
As one of the simplest models exhibiting nonreciprocal body motion, Najafi and Golestanian proposed a model 
of a three-sphere microswimmer~\cite{Najafi04,Golestanian08}, in which three in-line spheres are linked by two arms
of varying lengths. 
Later, such a three-sphere microswimmer has been experimentally realized~\cite{Leoni09,Grosjean16,Grosjean18}.
Various extensions of the original three-sphere microswimmer model were considered and are summarized
in Ref.~\cite{Yasuda23}.
Among them, we focus on the two stochastic microswimmers, i.e., thermal microswimmer in which the 
spheres have different temperatures~\cite{Hosaka17,Sou19,Sou21} and odd microswimmer in which the two springs 
(rather than arms) have odd elasticity~\cite{Yasuda21,Kobayashi23}.

The concept of odd elasticity was proposed by Scheibner \textit{et al.}\ to account for 
nonreciprocal interactions in active systems~\cite{Scheibner20,Fruchart23}.
They showed that the odd component of the elastic constant matrix quantifies the amount of work extracted along 
quasistatic deformation cycles.
Generalized odd elasticity exists not only in elastic materials but also in generic micromachines, such as 
molecular motors and catalytic enzymes that exhibit nonequilibrium steady state 
dynamics~\cite{Yasuda21catalytic,YasudaOM22,YasudaTCF22,KobayashiODD23,Ishimoto22,Ishimoto23}.
Since the thermal microswimmer model was proposed before the work by Scheibner \textit{et al.}, it is 
necessary to understand how it can be quantitatively characterized in terms of effective odd elasticity and 
to further compare it with the odd microswimmer.

For this purpose, we discuss various quantities obtained from the two stochastic microswimmer models 
and particularly focus on their time-correlation functions. 
This is because the antisymmetric part of the cross-correlation function can exist for nonequilibrium 
micromachines when the time-reversal symmetry is broken~\cite{YasudaTCF22,KobayashiODD23}.
In order to have a proper comparison between the two models, we generalize the odd microswimmer model 
to have different even elastic constants.  
On the other hand, we mostly neglect hydrodynamic interactions between the different spheres except 
when we calculate the average velocity.
In the absence of hydrodynamic interactions, the considered models do not exhibit any directed locomotion 
but undergo Brownian motion~\cite{Sou19,Sou21}.
Hence, we use the word ``micromachine" instead of ``microswimmer" throughout this paper.

After explaining the thermal three-sphere micromachine and the odd three-sphere micromachine, 
we show the average velocity, entropy production rate, 
extension-extension and velocity-velocity time correlation functions, and the corresponding Green-Kubo 
relations. 
The antisymmetric parts of the cross-correlation functions are explicitly obtained for the two micromachines,
reflecting the degree of broken time-reversal symmetry.  
For the odd micromachine, the time-correlation functions exhibit oscillatory behaviors when the odd 
elasticity is large enough.
Comparing the respective Green-Kubo relations for the two models, we show that the effective odd elasticity 
of the thermal microswimmer is proportional to the temperature difference between the two edge spheres.
Such a temperature difference causes a heat flow inside the micromachine, which can be quantified by 
the entropy production rate.

In Sec.~\ref{Sec:Odd}, we review the model of the odd three-sphere micromachine and show 
its statistical properties as mentioned above. 
In Sec.~\ref{Sec:Ther}, we perform a similar analysis for the thermal three-sphere micromachine.
In Sec.~\ref{Sec:Oddelasticity}, we compare the two stochastic micromachines and discuss the 
effective odd elasticity of the thermal micromachine. 
A summary and some further discussion are given in Sec.~\ref{Sec:Dis}.
In the Appendices, we show the derivations of the various time-correlation functions
for the two models.

\section{Odd three-sphere micromachine}
\label{Sec:Odd}

\begin{figure}[tb]
\begin{center}
\includegraphics[scale=0.38]{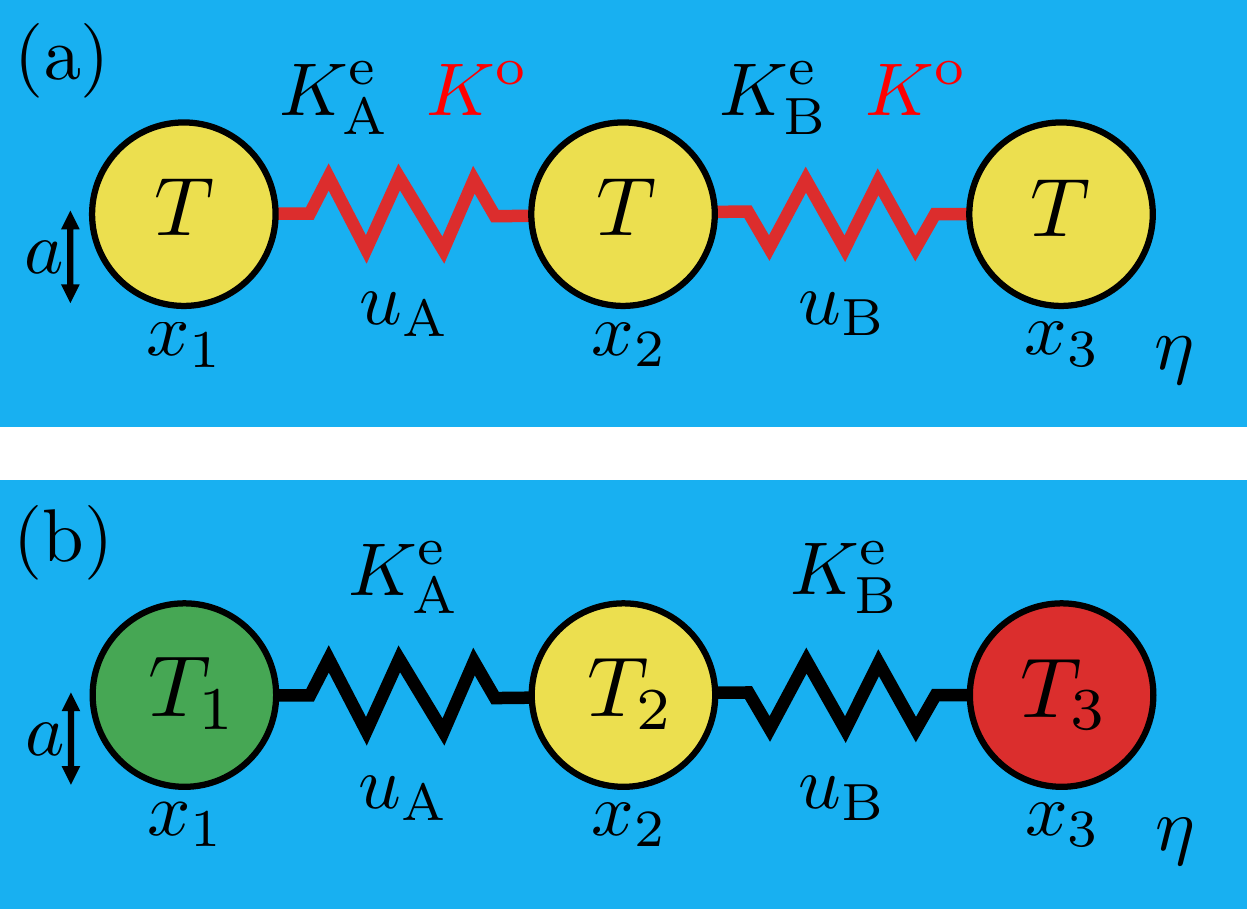}
\end{center}
\caption{
Two stochastic three-sphere micromachines consisting of three spheres with a radius $a$ and two springs 
with a natural length $\ell$.
These micromachines are immersed in a fluid with viscosity $\eta$.  
The positions of the spheres are denoted by $x_i$ ($i=1,2,3$) in a one-dimensional coordinate, and the spring 
extensions with respect to $\ell$ are denoted by $u_{\alpha}$ ($\alpha={\rm A}, {\rm B}$). 
(a) Odd three-sphere micromachine in which the two springs have the odd elastic constant $K^{\rm o}$
in addition to the even elastic constants $K^{\rm e}_{\rm A}$ and $K^{\rm e}_{\rm B}$.
The temperature of all the spheres is $T$. 
The ratios are defined by   
$\kappa=K_{\rm B}^{\rm e}/K_{\rm A}^{\rm e}$ and $\lambda=K^{\rm o}/K_{\rm A}^{\rm e}$.
(b) Thermal three-sphere micromachine in which the two springs have only the two elastic constants 
$K^{\rm e}_{\rm A}$ and $K^{\rm e}_{\rm B}$. 
However, the three spheres are in equilibrium with independent heat baths at different temperatures $T_i$ ($i=1,2,3$).
}
\label{model1}
\end{figure}

\subsection{Model}

We first describe the model of the odd micromachine that is slightly extended from our 
previous work~\cite{Yasuda21,Kobayashi23}. 
As schematically shown in Fig.~\ref{model1}(a), this model consists of three spheres of radius $a$ positioned
along a one-dimensional coordinate system, denoted by $x_i$ ($i=1,2,3$). 
These three spheres are connected by two springs that exhibit both even and odd elasticities. 
We denote the two spring extensions by $u_{\rm A}=x_2-x_1-\ell$ and $u_{\rm B}=x_3-x_2-\ell$, 
where $\ell$ is the constant natural length. 
Then the forces $F_{\rm A}$ and $F_{\rm B}$ conjugate to $u_{\rm A}$ and $u_{\rm B}$,
respectively, are given by $F_{\alpha}=-K_{\alpha\beta}u_{\beta}$ ($\alpha,\beta={\rm A},{\rm B}$).
For an odd elastic spring, the elastic constant matrix $K_{\alpha\beta}$ is given by
\begin{align} 
\mathbf{K}
=\left(                 
\begin{array}{cc}   
K_{\rm A}^{\rm e} & K^{\rm o} \vspace{1ex} 
\\ 
-K^{\rm o} & K_{\rm B}^{\rm e} \\
\end{array}
\right).
\label{ECK}      
\end{align}
In the two-dimensional configuration space, $K_{\rm A}^{\rm e}$ and $K_{\rm B}^{\rm e}$ are the positive 
even elastic constants, while $K^{\rm o}$ is the odd elastic constant that can take both positive and 
negative values.
In our previous study, we only studied the case of $K_{\rm A}^{\rm e}=K_{\rm B}^{\rm e}$~\cite{Yasuda21,Kobayashi23}.
The forces $f_i$ acting on each sphere are given by $f_1=-F_{\rm A}$,
$f_2=F_{\rm A}-F_{\rm B}$, and $f_3=F_{\rm B}$. 
Notice that these forces automatically satisfy the force-free condition, i.e., $f_1+f_2+f_3=0$~\cite{Yasuda17a}.

The above odd micromachine is immersed in a fluid of shear viscosity $\eta$ and temperature $T$.
The Langevin equation of each sphere is given by~\cite{DoiBook}
\begin{align} 
\dot{x}_i=M_{ij}f_j+\xi_{i},
\label{LEq}
\end{align}
where $\dot{x}_i=dx_{i}/dt$ and $M_{ij}$ is the hydrodynamic mobility coefficient matrix~\cite{Najafi04,Golestanian08}
\begin{align}
M_{ij} =\begin{cases}
1/(6\pi\eta a) & (i=j),\vspace{1ex} \\
1/(4\pi\eta |x_i-x_j\vert) & (i\neq j).
\end{cases}
\label{HDMC}
\end{align}
In Eq.~(\ref{LEq}), the Gaussian white-noise sources $\xi_{i}$ have zero mean $\langle \xi_i(t)\rangle=0$, 
and their correlations satisfy the fluctuation-dissipation theorem~\cite{DoiBook}
\begin{align} 
\langle \xi_i(t)\xi_j(t')\rangle=2k_{\rm B}TM_{ij}\delta(t-t'),
\label{FDT} 
\end{align} 
where $k_{\rm B}$ is the Boltzmann constant.
This condition assures that the surrounding fluid is in thermal equilibrium. 
Since the noise amplitudes depend on the particle positions, we use the It\^{o} interpretation in 
Eq.~(\ref{LEq})~\cite{Kobayashi23}.

\subsection{Average velocity}

It is convenient to introduce the characteristic time scale $\tau=6\pi\eta a/K_{\rm A}^{\rm e}$
describing the spring relaxation time.
We define the ratios between the elastic constants by 
$\kappa=K_{\rm B}^{\rm e}/K_{\rm A}^{\rm e}$ and $\lambda=K^{\rm o}/K_{\rm A}^{\rm e}$.
In the following analysis, we assume $u_\alpha\ll \ell$ and $a \ll \ell$, and focus 
on the leading-order contribution. 
The instantaneous total velocity of the micromachine is simply given by $V=(\dot{x}_1+\dot{x}_2+\dot{x}_3)/3$.
Using Eqs.~(\ref{ECK})-(\ref{HDMC}) and taking the statistical average, we obtain~\cite{Yasuda21,Kobayashi23}
\begin{align} 
\langle V\rangle &= \frac{a}{8\ell^2\tau} \Big[ (-1+3\lambda)\langle u_{\rm A}^2\rangle
+(\kappa+3\lambda)\langle u_{\rm B}^2\rangle \nonumber\\
&  +[3(1-\kappa)-2\lambda]\langle u_{\rm A}u_{\rm B}\rangle \Big],
\label{AveVO1}
\end{align}
where we have used $\langle u_\alpha \rangle=0$.

The equal-time correlation functions $\langle u_\alpha u_\beta \rangle$ appearing in Eq.~(\ref{AveVO1}) 
can be obtained from the reduced Langevin equations for
$\dot{u}_{\rm A}=\dot{x}_2-\dot{x}_1$ and $\dot{u}_{\rm B}=\dot{x}_3-\dot{x}_2$ as
\begin{align} 
\dot{u}_{\alpha}=\Gamma_{\alpha\beta}u_{\beta}+\Xi_{\alpha}, 
\label{RLEq}
\end{align}
where the friction matrix $\Gamma_{\alpha\beta}$ and the noise vector $\Xi_{\alpha}$ are given by 
\begin{align} 
\mathbf{\Gamma}
=-\frac{1}{\tau}
\left(                 
\begin{array}{cc}   
2+\lambda & -\kappa+2\lambda \\ 
-1-2\lambda & 2\kappa-\lambda \\
\end{array}
\right),
\quad
\mathbf{\Xi}
=\left(                 
\begin{array}{cc}   
\xi_2-\xi_1 \\ 
\xi_3-\xi_2 \\
\end{array}
\right).
\end{align}
Here, under the assumption $a\ll \ell$, we have neglected the hydrodynamic interactions acting 
between different spheres, i.e., $M_{ij} \approx 0$ when $i \neq j $ in Eq.~(\ref{HDMC}).
Such an approximation is justified when the sphere size is much smaller than the average spring length.
In general, the friction matrix $\Gamma_{\alpha\beta}$ is asymmetric, i.e., 
$\Gamma_{\rm{AB}}\neq \Gamma_{\rm{BA}}$. 
Such an asymmetry arises both from the different even elastic constants ($\kappa \neq 1$) and from the 
presence of the odd elasticity ($\lambda \neq 0$).
In our previous works~\cite{Yasuda21,Kobayashi23}, we focused on the case 
when $\kappa=1$ and $\lambda \neq 0$.

To deal with the above Langevin equations, we introduce the bilateral Fourier transform of a function $f(t)$ 
as $f(\omega)=\int_{-\infty}^{\infty}dt \,f(t)e^{-{\rm i}\omega t}$
and its inverse transform as $f(t)=\int_{-\infty}^{\infty}(d\omega/2\pi) \, f(\omega)e^{{\rm i}\omega t}$.
Solving Eq.~(\ref{RLEq}) in the frequency domain, we obtain
\begin{widetext}
\begin{align}
u_{\rm A}(\omega) & = \frac{\left[(2\kappa-\lambda)+{\rm i}\omega\tau\right]\xi_1(\omega)
-\left[(\kappa+\lambda)+{\rm i}\omega\tau\right]\xi_2(\omega)-(\kappa-2\lambda)\xi_3(\omega)}
{-3(\kappa+\lambda^2)-2(1+\kappa){\rm i}\omega\tau+(\omega\tau)^2}\tau,
\label{FreqU_A} \\ 
u_{\rm B}(\omega) &= \frac{(1+2\lambda)\xi_1(\omega)
+\left[(1-\lambda)+{\rm i}\omega\tau\right]\xi_2(\omega)-\left[(2+\lambda)+{\rm i}\omega\tau\right]\xi_3(\omega)}
{-3(\kappa+\lambda^2)-2(1+\kappa){\rm i}\omega\tau+(\omega\tau)^2}\tau.
\label{FreqU_B} 
\end{align}
\end{widetext}
Using Eqs.~(\ref{FreqU_A}) and (\ref{FreqU_B}), one can calculate the correlation functions 
$\langle u_{\rm \alpha}(\omega)u_{\rm \beta}(\omega')\rangle$.
Then, the equal-time correlation functions are obtained as
\begin{align}
\langle u_{\rm A}^{2}\rangle&=\int_{-\infty}^{\infty}\frac{d\omega}{2\pi}
\int_{-\infty}^{\infty}\frac{d\omega'}{2\pi}
\langle u_{\rm A}(\omega)u_{\rm A}(\omega')\rangle \nonumber \\
&=\frac{k_{\rm B}T(\kappa+\kappa^2-\kappa\lambda+2\lambda^2)}{K_{\rm A}^{\rm e}(1+\kappa)(\kappa+\lambda^2)},
\label{ETc-AA} \\ 
\langle u_{\rm B}^{2}\rangle
&=\frac{k_{\rm B}T(1+\kappa+\lambda+2\lambda^2)}{K_{\rm A}^{\rm e}(1+\kappa)(\kappa+\lambda^2)}, 
\label{ETc-BB} \\ 
\langle u_{\rm A}u_{\rm B}\rangle
&=- \frac{k_{\rm B}T[(1-\kappa)\lambda+\lambda^2]}{K_{\rm A}^{\rm e}(1+\kappa)(\kappa+\lambda^2)}.
\label{ETc-AB} 
\end{align}
Here, we have neglected the cross-correlations of the noise, $\langle \xi_{i}\xi_{j}\rangle$ with $i\neq j$,  
because they only contribute to the higher order terms in $a/\ell$.
When $\lambda=0$, the above expressions reduce to 
$\langle u_{\rm A}^{2}\rangle= k_{\rm B}T/K_{\rm A}^{\rm e}$, 
$\langle u_{\rm B}^{2}\rangle = k_{\rm B}T/K_{\rm B}^{\rm e}$,  
and $\langle u_{\rm A}u_{\rm B}\rangle=0$, reproducing the well-known thermal equilibrium case.

Substituting the equal-time correlations in Eqs.~(\ref{ETc-AA})-(\ref{ETc-AB}) into Eq.~(\ref{AveVO1}), we 
obtain the steady state average velocity of the odd micromachine as
\begin{align} 
\langle V\rangle=\frac{7k_{\rm B}T\lambda}{24\pi\eta\ell^2(1+\kappa)},
\label{AveV02}
\end{align}
which reduces to the previous result when $\kappa=1$~\cite{Yasuda21,Kobayashi23}.
Recalling $\lambda=K^{\rm o}/K_{\rm A}^{\rm e}$, we see that $\langle V\rangle$ is proportional 
to the odd elastic constant $K^{\rm o}$ whose sign determines the swimming direction.
Since $\langle V\rangle$ is also proportional to $k_{\rm B}T$, thermal fluctuations are responsible for 
the locomotion of the odd micromachine.

\subsection{Entropy production rate}

Next, we calculate the steady state average entropy production rate $\langle \dot{\sigma}\rangle$
of the odd three-sphere micromachine, which is given by 
\begin{align} 
\langle \dot{\sigma}\rangle=-\sum_{i=1}^3 \frac{\langle \dot{Q}_i \rangle}{T}.
\label{EPROM}
\end{align}
According to the framework of stochastic energetics established by Sekimoto~\cite{SekimotoBook}, 
the time-derivative of the heat gained by the $i$-th sphere is 
\begin{align} 
\dot{Q}_i =6\pi\eta a(-\dot{x}_i+\xi_i)\circ\dot{x}_i,
\label{HFPS}
\end{align}
where $\dot{x}_i$ and $\xi_i$ are given by Eq.~(\ref{LEq}) and 
$\circ$ indicates the Stratonovich product (the summation over $i$ is not taken in this equation).
Then, the lowest-order average heat flows become 
\begin{align}
{\langle \dot{Q}_1 \rangle}&=\frac{k_{\rm B}T\lambda(1-2\lambda)}{\tau(1+\kappa)}, 
\label{AHFO1}\\ 
{\langle \dot{Q}_2 \rangle}&=\frac{k_{\rm B}T\lambda(-1+\kappa-2\lambda)}{\tau(1+\kappa)}, 
\label{AHFO2}\\ 
{\langle \dot{Q}_3\rangle}&=-\frac{k_{\rm B}T\lambda(\kappa+2\lambda)}{\tau(1+\kappa)},
\label{AHFO3} 
\end{align}
which all vanish when $\lambda=0$. 
Substituting Eqs.~(\ref{AHFO1})-(\ref{AHFO3}) into Eq.~(\ref{EPROM}), we obtain the average entropy 
production rate of the odd micromachine as  
\begin{align} 
\langle \dot{\sigma}\rangle=\frac{6k_{\rm B}\lambda^2}{\tau(1+\kappa)},
\label{AEPRO1}
\end{align}
which is obviously non-negative, $\langle \dot{\sigma}\rangle \ge 0$, and vanishes only when $\lambda=0$.
This result is in accordance with the second law for nonequilibrium steady states~\cite{Freitas22}.
Equation~(\ref{AEPRO1}) is also consistent with the fact that the entropy production 
rate should be independent of the sign of the odd elastic constant and its lowest-order contribution 
is proportional to $\lambda^2$.

In our previous work with $\kappa=1$~\cite{Yasuda21}, we obtained the entropy production rate by 
using the combination of the friction matrix and the covariant matrix of the Gaussian distribution 
function~\cite{Weiss03,Weiss07}.
We also showed that the entropy production rate coincides with the power (work per unit time) 
of the odd micromachine. 
Hence, all the extracted work due to odd elasticity is converted into entropy production in the 
steady state.

\begin{figure*}[tb]
\centering
\includegraphics[scale=0.58]{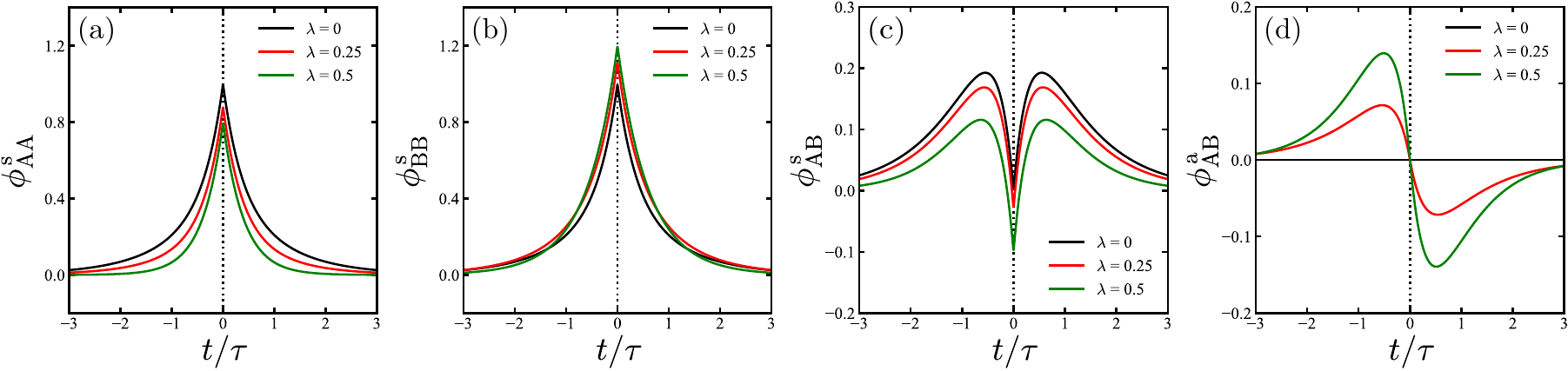}
\caption{Plots of the scaled extension-extension correlation functions 
(a) $\phi_{\rm{AA}}^{\rm s}$, (b) $\phi_{\rm{BB}}^{\rm s}$, (c) $\phi_{\rm{AB}}^{\rm s}$, and 
(d) $\phi_{\rm{AB}}^{\rm a}$ as a function of $t/\tau$ for $\kappa=1$.
The dimensionless odd elasticity is chosen as $\lambda=0$ (black), $0.25$ (red), and 
$0.5$
(green).
For these parameters, $\mu=\sqrt{3\lambda^2-(1-\kappa+\kappa^2)}$ is purely imaginary, and 
the correlation functions do not exhibit any oscillatory behaviors. 
}
\label{Fig:DTCFOim}
\end{figure*}

\begin{figure*}[tb]
\centering
\includegraphics[scale=0.58]{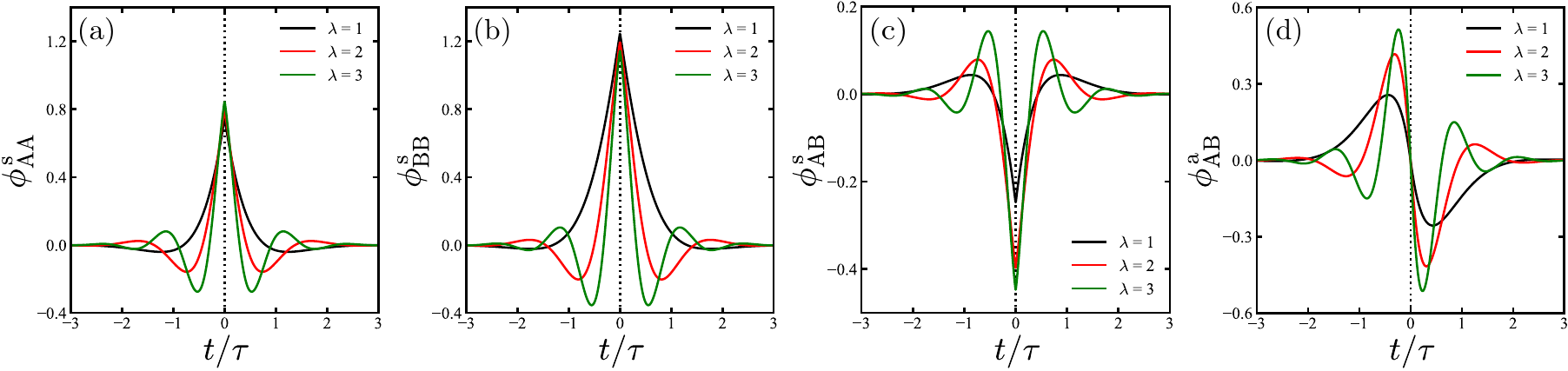}
\caption{Plots of the scaled extension-extension correlation functions 
(a) $\phi_{\rm{AA}}^{\rm s}$, (b) $\phi_{\rm{BB}}^{\rm s}$, (c) $\phi_{\rm{AB}}^{\rm s}$, and 
(d) $\phi_{\rm{AB}}^{\rm a}$ as a function of $t/\tau$ for $\kappa=1$.
The dimensionless odd elasticity is chosen as $\lambda=1$ (black), $2$ (red), and $3$ (green).
For these parameters, $\mu=\sqrt{3\lambda^2-(1-\kappa+\kappa^2)}$ is real and
the correlation functions exhibit oscillatory behaviors. 
}
\label{Fig:DTCFOre}
\end{figure*}

\subsection{Time-correlation functions}

Next, we calculate the time-correlation functions of the odd micromachine in the steady state.
Using Eqs.~(\ref{FreqU_A}) and (\ref{FreqU_B}), we perform inverse Fourier transform and obtain 
the extension-extension time-correlation function matrix $\phi_{\alpha\beta}(t)$ defined by 
\begin{align}
\langle u_{\alpha}(t)u_{\beta}(0)\rangle=\frac{k_{\rm B} T}{K_{\rm A}^{\rm e}}\phi_{\alpha\beta}(t),
\end{align}
where $\phi_{\alpha\beta}$ is dimensionless. 
Generally, one can decompose the time-correlation functions into the symmetric and antisymmetric  
parts as~\cite{YasudaTCF22,KobayashiODD23} 
\begin{align}
\phi_{\alpha\beta}(t) =\phi_{\alpha\beta}^\mathrm{s}(t)+\phi_{\alpha\beta}^\mathrm{a}(t),
\label{decomposephi}
\end{align}
where $\phi_{\alpha\beta}^\mathrm s(t)=\phi_{\beta\alpha}^\mathrm s(t)$ and 
$\phi_{\alpha\beta}^\mathrm a(t)=-\phi_{\beta\alpha}^\mathrm a(t)$.
When $\alpha=\beta$ (auto-correlation), only the symmetric part is allowed due to the time-translational 
invariance of the steady state and hence the antisymmetric parts should vanish, i.e., 
$\phi_{\rm{AA}}^{\rm a} =\phi_{\rm{BB}}^{\rm a}=0$.
When $\alpha\neq\beta$ (cross-correlation) and if the system is in thermal equilibrium, the 
antisymmetric part of the cross-correlation $\phi_{\rm{AB}}^{\rm a}$ should also vanish due to the 
time-reversal symmetry~\cite{DoiBook}.
In nonequilibrium situations, however, $\phi_{\rm{AB}}^{\rm a}$ can exist because the time-reversal 
symmetry is generally broken~\cite{YasudaTCF22,KobayashiODD23}.
In other words, $\phi_{\rm{AB}}^{\rm a}$ quantifies the degree of nonequilibrium in stochastic 
micromachines.

The inverse Fourier transform of the correlation function can be straightforwardly performed by 
employing the residue theorem (see Appendix~\ref{App:A} for the details).
After some calculation, we obtain the time-dependent extension-extension 
correlation functions as
\begin{align}
\phi_{\rm{AA}}^{\rm s}(t)
& =\frac{\kappa+\kappa^2-\kappa\lambda+2\lambda^2}{(1+\kappa)(\kappa+\lambda^2)}
\cos\left(\mu t/\tau \right)e^{-(1+\kappa)\vert t \vert/\tau}
\nonumber\\
&-\frac{\kappa(1-\kappa+\lambda)}{(\kappa+\lambda^2)\mu}
\sin \left(\mu \vert t \vert/\tau \right)
e^{-(1+\kappa) \vert t \vert/\tau},
\label{DTCFO_AA}\\ 
\phi_{\rm{BB}}^{\rm s}(t)
& =\frac{1+\kappa+\lambda+2\lambda^2}{(1+\kappa)(\kappa+\lambda^2)}
\cos \left (\mu t/\tau \right)
e^{-(1+\kappa)\vert t \vert/\tau}
\nonumber\\
&+\frac{1-\kappa+\lambda}{(\kappa+\lambda^2)\mu}
\sin \left(\mu \vert t \vert/\tau \right)e^{-(1+\kappa) \vert t \vert/\tau},
\label{DTCFO_BB}\\ 
\phi_{\rm{AB}}^{\rm s}(t)
&=-\frac{(1-\kappa)\lambda+\lambda^2}{(1+\kappa)(\kappa+\lambda^2)}
\cos\left(\mu t/\tau \right)
e^{-(1+\kappa) \vert t \vert/\tau}
\nonumber\\
& + \frac{\kappa-(1-\kappa)\lambda}{(\kappa+\lambda^2)\mu}
\sin \left(\mu \vert t \vert/\tau \right)
e^{-(1+\kappa) \vert t \vert/\tau},
\label{DTCFO_AB1}\\ 
\phi_{\rm{AB}}^{\rm a}(t)
& =-\frac{3\lambda}{(1+\kappa)\mu}
\sin \left(\mu t/\tau \right)
e^{-(1+\kappa)\vert t \vert/\tau},
\label{DTCFO_AB2} 
\end{align}
where we have used the notation $\mu=\sqrt{3\lambda^2-(1-\kappa+\kappa^2)}$.
When $\lambda^2 < (1-\kappa+\kappa^2)/3$, $\mu$ is purely imaginary so that 
``$\cos$" and ``$\sin$" functions should be regarded as ``$\cosh$" and ``$\sinh$" functions, respectively.
The equal-time correlations in Eqs.~(\ref{ETc-AA})-(\ref{ETc-AB}) can be recovered by setting $t=0$ 
in the above equations.

As mentioned before, the auto-correlations have only the symmetric parts $\phi_{\rm{AA}}^{\rm s}$
and $\phi_{\rm{BB}}^{\rm s}$, whereas the cross-correlation contains both the 
symmetric part $\phi_{\rm{AB}}^{\rm s}$ and the antisymmetric part $\phi_{\rm{AB}}^{\rm a}$.
However, $\phi_{\rm{AB}}^{\rm a}$ exists only for nonzero $\lambda$ and it measures the degree 
of the broken time-reversal symmetry.
Obviously, the symmetric parts are even functions in time, and the antisymmetric part is an odd function in time.
In Fig.~\ref{Fig:DTCFOim}, we plot (a) $\phi_{\rm{AA}}^{\rm s}$, (b) $\phi_{\rm{BB}}^{\rm s}$, 
(c) $\phi_{\rm{AB}}^{\rm s}$, and (d) $\phi_{\rm{AB}}^{\rm a}$ as a function of $t/\tau$ 
when $\kappa=1$ and $\lambda=0, 0.25, 0.5$ (when $\mu$ is imaginary). 
In Fig.~\ref{Fig:DTCFOre}, we plot the same functions  
when $\kappa=1$ and $\lambda=1, 2, 3$ (when $\mu$ is real). 
In Figs.~\ref{Fig:DTCFOim}(a) and (b), the auto-correlation functions decay monotonically, whereas 
the symmetric part of the cross-correlation function exhibits maximum values and even takes negative 
values in Figs.~\ref{Fig:DTCFOim}(c).
As we increase the odd elasticity $\lambda$ in Fig.~\ref{Fig:DTCFOre}, the oscillatory behaviors of 
the time-correlation functions become pronounced.
The symmetric part of the cross-correlation function $\phi_{\rm{AB}}^{\rm s}$ takes 
maximum values at around $t/\tau \approx 1$ (rather than at $t=0$) because the time needed for 
the communication between the two springs is the order of the spring relaxation time $\tau$.
In other words, $\tau$ can be regarded as an effective viscoelastic time scale that describes the retarded correlation.

The velocity-velocity correlation function matrix $\psi_{\alpha\beta}(t)$ can be obtained by taking 
the second time derivative of the extension-extension correlation functions as~\cite{DoiBook}
\begin{align} 
\langle \dot{u}_{\alpha}(t)\dot{u}_{\beta}(0)\rangle=
-\frac{\partial ^2}{\partial t^2}
\langle u_{\alpha}(t)u_{\beta}(0)\rangle = 
\frac{D}{\tau} \psi_{\alpha\beta}(t),
\label{RVDTCFO}
\end{align}
where we have introduced the diffusion coefficient of a sphere $D=k_{\rm B}T/(6\pi\eta a)$. 
Alternatively, one can also use Eq.~(\ref{RLEq}) to calculate $\langle \dot{u}_{\alpha}(t)\dot{u}_{\beta}(0)\rangle$ 
directly, as shown in Appendix~\ref{App:C}.
Similar to Eq.~(\ref{decomposephi}), $\psi_{\alpha\beta}(t)$ can be decomposed into the symmetric and antisymmetric
parts as 
$\psi_{\alpha\beta}(t) =\psi_{\alpha\beta}^\mathrm s(t)+\psi_{\alpha\beta}^\mathrm a(t)$.
Following the same argument as before, we have $\psi_{\rm{AA}}^{\rm a} =\psi_{\rm{BB}}^{\rm a}=0$
for the auto-correlation functions.
However, the antisymmetric part of the cross-correlation $\psi_{\rm{AB}}^{\rm a}$ exists when $\lambda$ is nonzero.

The explicit expressions of the velocity-velocity correlation functions and their plots are given in Appendix~\ref{App:C}.
We note that the symmetric part of the velocity-velocity correlation functions has a sharp peak at $t=0$, 
which arises from the noise term in the Langevin equation.
For very short time, $t \ll \tau$, the velocity-velocity correlation functions are approximated as~\cite{DoiBook} 
\begin{align} 
\langle \dot{u}_{\rm A}(t)\dot{u}_{\rm A}(0)\rangle & = \langle \dot{u}_{\rm B}(t)\dot{u}_{\rm B}(0)\rangle \approx 4D \delta(t), 
\label{shorttimeAA}
\\ 
\langle \dot{u}_{\rm A}(t)\dot{u}_{\rm B}(0)\rangle & =\langle \dot{u}_{\rm B}(t)\dot{u}_{\rm A}(0)\rangle \approx -2D \delta(t).
\label{shorttimeAB}
\end{align}

\subsection{Green-Kubo relations}

At equilibrium, the time integral of the velocity-velocity correlation function for a free particle 
gives the diffusion coefficient, known as one of the Green-Kubo relations~\cite{Zwanzig}.
Recently, Han \textit{et al.}\ argued that an equilibrium-like Green-Kubo relation holds near the 
steady state of an isotropic active fluid if the activated and fluctuating degrees of freedom are 
statistically decoupled~\cite{Han21}.
Epstein and Mandadapu obtained the Green-Kubo relation for the odd viscosity~\cite{Avron98}, 
which explicitly violates the time-reversal symmetry at the level of the steady state stress 
fluctuations~\cite{Epstein20,Hargus20}.

For the odd micromachine, we perform the time integral of the velocity-velocity correlation functions 
given in Appendix~\ref{App:C}.
From the auto-correlation functions in Eqs.~(\ref{SVTCFO_AA}) and (\ref{SVTCFO_BB}), we obtain
\begin{align}
\int_{0}^{\infty} dt\, \langle \dot{u}_{\rm A}(t)\dot{u}_{\rm A}(0)\rangle
=\int_{0}^{\infty} dt\, \langle  \dot{u}_{\rm B}(t)\dot{u}_{\rm B}(0)\rangle=0.
\label{GKRO_3}
\end{align}
Here, the time integral of Eq.~(\ref{shorttimeAA}) gives $2D$ which is canceled by the rest of the terms
in Eqs.~(\ref{SVTCFO_AA}) and (\ref{SVTCFO_BB}).
The above result with the vanishing integrals in Eq.~(\ref{GKRO_3}) is reasonable because the particles 
are connected by the springs~\cite{DoiBook}.

For the cross-correlation functions in Eqs.~(\ref{VTCFO_AB1})-(\ref{VTCFO_AB2}), on the other hand, 
we have 
\begin{align}
\int_{0}^{\infty} dt\, 
\left[ \langle \dot{u}_{\rm A}(t)\dot{u}_{\rm B}(0)\rangle - 
\langle \dot{u}_{\rm B}(t)\dot{u}_{\rm A}(0)\rangle \right]
= -\frac{6\lambda D}{1+\kappa}.
\label{GKRO_5} 
\end{align}
In this calculation, only the antisymmetric part of the correlation function $\psi_{\rm AB}^\mathrm a$ remains. 
Equation~(\ref{GKRO_5}) explicitly demonstrates that the broken time-reversal symmetry is caused 
by the finite odd elasticity $\lambda$. 
The above result also suggests that one can estimate the odd elasticity of a micromachine by measuring 
the cross-correlation functions~\cite{KobayashiODD23}.

\section{Thermal three-sphere micromachine}
\label{Sec:Ther}

\subsection{Model}

In this section, we discuss another type of stochastic micromachine that is also purely driven by 
thermal fluctuation but without any explicit odd elasticity. 
Here, the key assumption is that the three spheres are in equilibrium with independent heat baths 
having different temperatures~\cite{Hosaka17,Sou19,Sou21}.
In this micromachine, the heat transfer occurs from a hotter sphere to a colder one, driving the 
whole system out of equilibrium.

As shown in Fig.~\ref{model1}(b), we consider a three-sphere micromachine in which the three spheres 
are in equilibrium with independent heat baths having different temperatures $T_i$
($i=1,2,3$)~\cite{Hosaka17,Sou19,Sou21}. 
On the other hand, the two springs have only even elasticity characterized by $K_{\rm A}^{\rm e}$ 
and $K_{\rm B}^{\rm e}$, whereas an explicit odd elastic constant does not exist.
The equation of motion of each sphere is the same as in Eq.~(\ref{LEq}), although the thermal noise 
$\xi_i$ has different statistical properties.   
For the thermal micromachine, we assume that the Gaussian white-noise sources $\xi_{i}$ have zero 
mean $\langle \xi_i(t)\rangle_{\rm t}=0$, and their correlations satisfy 
\begin{align} 
\langle \xi_i(t)\xi_j(t')\rangle_{\rm t}=2 D_{ij}\delta(t-t').
\label{FDT2} 
\end{align}
Here, $D_{ij}$ is the mutual diffusion coefficient matrix given by
\begin{align}
D_{ij} =\begin{cases}
k_{\rm B}T_i/(6\pi\eta a) & (i=j),\vspace{1ex} \\
k_{\rm B}\Theta(T_i,T_j)/(4\pi\eta |x_i-x_j\vert) & (i\neq j),
\end{cases}
\label{MDCT}
\end{align}
where $\Theta(T_i,T_j)$ is a function of $T_i$ and $T_j$. 
The effective temperature $\Theta$ can be the mobility-weighted average~\cite{Grosberg15}, which 
in the present case is simply given by $\Theta(T_i,T_j)=(T_i+T_j)/2$ because all the spheres have the 
same size. 
However, its explicit functional form is not needed here, and we only require that $\Theta$
should satisfy an appropriate fluctuation-dissipation theorem in thermal equilibrium.
Such a simplification is justified because we only consider the limit of $a \ll \ell$.

\subsection{Average velocity}

The average velocity of the thermal micromachine can be obtained similarly as before by using 
Eq.~(\ref{AveVO1}) with $\lambda=0$ because the odd elasticity does not exist in this model. 
In our previous paper, the average velocity was obtained as~\cite{Hosaka17} 
\begin{align} 
\langle V\rangle_{\rm t} =\frac{k_{\rm B}T_2\left[(2-5\kappa)\theta_1-7(1-\kappa)+(5-2\kappa)\theta_3\right]}
{144\pi\eta\ell^2(1+\kappa)},
\label{TAveV1}
\end{align}
where $\kappa=K_{\rm B}^{\rm e}/K_{\rm A}^{\rm e}$ as before and we have defined the 
temperature ratios by $\theta_1=T_1/T_2$ and $\theta_3=T_3/T_2$.
When the three temperatures are identical, $T_1=T_2=T_3$ or $\theta_1=\theta_3=1$, the velocity vanishes 
identically, $\langle V\rangle_{\rm t}=0$. 
This means that the temperature difference among the spheres causes the locomotion of the 
thermal three-sphere micromachine.

When the two springs are equivalent and $\kappa=1$, $\langle V \rangle_{\rm t}$ does not depend on 
$T_2$ and is proportional to the temperature difference $T_3-T_1$.
In this case, the average velocity vanishes when $T_1=T_3$ even though $T_1$ and $T_3$ are
different from $T_2$.
This result is reasonable because such a thermal micromachine has a perfect head-tail symmetry 
when $\kappa=1$ and $T_1=T_3$, and it cannot acquire any directed motion due to thermal fluctuations.

\subsection{Entropy production rate}

The average entropy production rate can also be calculated similarly to the previous section. 
For the thermal micromachine, the lowest-order average heat flows were obtained as~\cite{Hosaka17} 
\begin{align}
\langle \dot{Q}_1 \rangle_{\rm t}
&=\frac{k_{\rm B}T_2[(3+2\kappa)\theta_1-(3+\kappa)-\kappa \theta_3]}{6\tau(1+\kappa)}, 
\label{EAHFT_1}\\ 
\langle \dot{Q}_2 \rangle_{\rm t}
&=\frac{k_{\rm B}T_2[-(3+\kappa)\theta_1+(3+2\kappa+3\kappa^2)-(\kappa+3\kappa^2)\theta_3]}{6\tau(1+\kappa)}, 
\label{EAHFT_2}\\ 
\langle \dot{Q}_3 \rangle_{\rm t}
&=\frac{k_{\rm B}T_2[-\kappa \theta_1-(\kappa+3\kappa^2)+(2\kappa+3\kappa^2)\theta_3 ]}{6\tau(1+\kappa)},
\label{EAHFT_3}
\end{align}
which all vanish when $T_1=T_2=T_3$ or $\theta_1=\theta_3=1$. 
It is worthwhile to note that the above heat flows satisfy 
$\langle \dot{Q}_1 \rangle_{\rm t} + \langle \dot{Q}_2 \rangle_{\rm t} + \langle \dot{Q}_3 \rangle_{\rm t}=0$
so that the total heat is conserved.

Using these results and keeping in mind the different temperatures of the spheres, the steady state average entropy 
production rate is given by 
\begin{align} 
\langle \dot{\sigma}\rangle_{\rm t}
= & -\sum_{i=1}^3 \frac{\langle \dot{Q}_i \rangle_{\rm t}}{T_i}  \nonumber\\
= &\frac{k_{\rm B}}{6\tau(1+\kappa) } \Biggl[
(3+\kappa) \left( \theta_1 + \frac{1}{\theta_1} \right) 
+ \kappa \left( \frac{\theta_1}{\theta_3} + \frac{\theta_3}{\theta_1} \right) 
 \nonumber\\
& + \kappa(1+3\kappa) \left( \theta_3 + \frac{1}{\theta_3} \right) -6(1+\kappa+\kappa^2) \Biggr].
\label{TAEPR1}
\end{align}
One can easily confirm the second law, $\langle \dot{\sigma}\rangle_{\rm t} \ge 0$, by using the 
inequality $A+B \ge 2 \sqrt{AB}$ for $A, B>0$ and $\langle \dot{\sigma}\rangle_{\rm t}$ 
vanishes when $\theta_1=\theta_3=1$.
When $\kappa=1$ and $\theta_1=\theta_3 \neq 1$ (or $T_1=T_3 \neq T_2$), however, notice that 
$\langle \dot{\sigma}\rangle_{\rm t}$ is nonzero even though $\langle V\rangle_{\rm t}=0$.
The above entropy production rate reduces to that in Ref.~\cite{Sou21} when $\kappa=1$.

\begin{figure*}[tb]
\centering
\includegraphics[scale=0.58]{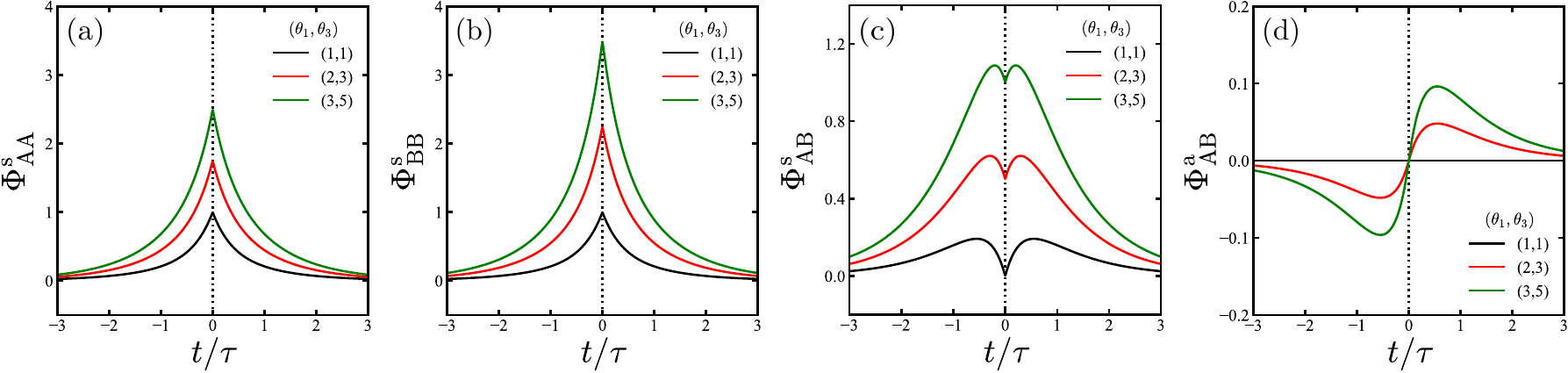}
\caption{Plots of the scaled extension-extension correlation functions 
(a) $\Phi_{\rm{AA}}^{\rm s}$, (b) $\Phi_{\rm{BB}}^{\rm s}$, (c) $\Phi_{\rm{AB}}^{\rm s}$, and 
(d) $\Phi_{\rm{AB}}^{\rm a}$ as a function of $t/\tau$ for $\kappa=1$.
The dimensionless temperature ratios between the spheres are chosen as 
$(\theta_1,\theta_3)=(1,1)$ (black), 
$(2,3)$ (red), and 
$(3,5)$ (green).}
\label{Fig:DTCFT}
\end{figure*}

\subsection{Time-correlation functions}

Next, we calculate the extension-extension time-correlation function matrix $\Phi_{\alpha\beta}$
for the thermal micromachine and decompose it into the symmetric and antisymmetric
parts as  
\begin{align}
\langle u_{\alpha}(t)u_{\beta}(0)\rangle_{\rm t} 
=\frac{k_{\rm B} T_2}{K_{\rm A}^{\rm e}} \Phi_{\alpha\beta}(t)
=\frac{k_{\rm B} T_2}{K_{\rm A}^{\rm e}}
\left[\Phi_{\alpha\beta}^\mathrm s(t)+\Phi_{\alpha\beta}^\mathrm a(t) \right].
\end{align}
The time-translational invariance requiers $\Phi_{\rm{AA}}^{\rm a} =\Phi_{\rm{BB}}^{\rm a}=0$ as before.
After repeating similar calculations 
(see Appendix~\ref{App:B} for the details), 
we obtain
\begin{align}
\Phi_{\rm AA}^\mathrm s(t)
& = \frac{(3+4\kappa)\theta_1+(3+\kappa)+\kappa \theta_3}{6(1+\kappa)} 
\cosh \left (\rho t/\tau \right)
e^{-(1+\kappa) \vert t \vert/\tau} 
\nonumber\\
& + \frac{(-3+4\kappa)\theta_1 -(3-\kappa)+\kappa \theta_3}{6\rho}
\sinh \left (\rho \vert t \vert/\tau \right)
e^{-(1+\kappa)\vert t\vert/\tau},
\label{TCFT_AA} \\
\Phi_{\rm BB}^\mathrm s(t)
&= \frac{\theta_1+(1+3\kappa)+(4+3\kappa)\theta_3}{6\kappa(1+\kappa)}
\cosh \left(\rho t/\tau \right)
e^{-(1+\kappa) \vert t \vert/\tau}
\nonumber\\
& + \frac{\theta_1 +(1-3\kappa)+(4-3\kappa)\theta_3}{6\kappa\rho}
\sinh \left(\rho \vert t \vert/\tau \right)
e^{-(1+\kappa) \vert t \vert/\tau},
\label{TCFT_BB} \\
\Phi_{\rm{AB}}^{\rm s}(t)
& = \frac{\theta_1-2+\theta_3}{3(1+\kappa)}
\cosh \left(\rho t/\tau \right)
e^{-(1+\kappa)\vert t \vert/\tau}
\nonumber \\
& + \frac{\theta_1+1+\theta_3}{3\rho}
\sinh \left(\rho \vert t \vert/\tau \right)
e^{-(1+\kappa)\vert t \vert/\tau},
\label{TCFT_AB^S} \\
\Phi_{\rm{AB}}^{\rm a}(t) 
& = \frac{-\theta_1+(1-\kappa)+\kappa \theta_3}
{2(1+\kappa)\rho}
\sinh \left(\rho t/\tau \right)
e^{-(1+\kappa) \vert t \vert/\tau},
\label{TCFT_AB^A}
\end{align}
where we have used the notation $\rho=\sqrt{1-\kappa+\kappa^2}$.
Since $1-\kappa+\kappa^2>0$, $\rho$ is real for any $\kappa$, and all the above time-correlation 
functions do not oscillate in time.  
In Eq.~(\ref{TCFT_AB^A}), we see that $\Phi_{\rm{AB}}^{\rm a}(t)$ vanishes when $\theta_1=\theta_3=1$.

Next, we calculate the velocity-velocity correlation function matrix  $\Psi_{\alpha\beta}(t)$ for the 
thermal micromachine as   
\begin{align} 
\langle \dot{u}_{\alpha}(t)\dot{u}_{\beta}(0)\rangle_{\rm t} 
= \frac{D_{22}}{\tau} \Psi_{\alpha\beta} (t) 
= \frac{D_{22}}{\tau} 
\left[\Psi_{\alpha\beta}^\mathrm s(t)+\Psi_{\alpha\beta}^\mathrm a(t) \right],
\label{thermRVDTCFO}
\end{align}
where $D_{22}=k_{\rm B}T_2/(6\pi\eta a)$ is the diffusion coefficient defined by using the 
middle sphere temperature $T_2$ [see Eq.~(\ref{MDCT})]. 
The explicit expressions of the velocity-velocity correlation functions and their plots are given in 
Appendix~\ref{App:D}.
Similar to the odd micromachine, the symmetric parts of the velocity-velocity correlation functions have a 
sharp peak at $t=0$. 
In the short time limit, $t \ll \tau$, the velocity-velocity correlation functions become 
\begin{align} 
\langle \dot{u}_{\rm A}(t)\dot{u}_{\rm A}(0)\rangle_{\rm t} & \approx 2 (\theta_1+1) D_{22} \delta(t),
\label{shorttimeAAth}
\\ 
\langle \dot{u}_{\rm B}(t)\dot{u}_{\rm B}(0)\rangle_{\rm t} & \approx 2 (1+\theta_3) D_{22} \delta(t),
\label{shorttimeBBth}
\\
\langle \dot{u}_{\rm A}(t)\dot{u}_{\rm B}(0)\rangle_{\rm t} & =\langle \dot{u}_{\rm B}(t)\dot{u}_{\rm A}(0)\rangle_{\rm t} 
\approx -2 D_{22} \delta(t).
\label{shorttimeABth}
\end{align}
These results coincide with Eqs.~(\ref{shorttimeAA}) and (\ref{shorttimeAB}) when $\theta_1 =\theta_3=1$ and 
if $D_{22}$ can be identified with $D$.

\subsection{Green-Kubo relations}

Having obtained the velocity-velocity correlation functions for the thermal micromachine, 
we now calculate the corresponding Green-Kubo relations as in the previous section. 
From the auto-correlation functions in Eqs.~(\ref{PsiAA}) and (\ref{PsiBB}), we have 
\begin{align}
\int_{0}^{\infty} dt\, \langle \dot{u}_{\rm A}(t)\dot{u}_{\rm A}(0)\rangle_{\rm t}
= \int_{0}^{\infty} dt\, \langle  \dot{u}_{\rm B}(t)\dot{u}_{\rm B}(0)\rangle_{\rm t}
=0,
\label{GKRO_3therm}
\end{align}
similar to Eq.~(\ref{GKRO_3}). 
From the cross-correlation functions in Eqs.~(\ref{PsiABeven}) and (\ref{PsiABodd}), 
however, we obtain 
\begin{align}
& \int_{0}^{\infty} dt\,    
\left[ \langle \dot{u}_{\rm A}(t)\dot{u}_{\rm B}(0)\rangle_{\rm t}
- \langle \dot{u}_{\rm B}(t)\dot{u}_{\rm A}(0)\rangle_{\rm t} \right]
\nonumber \\
& = -\frac{[\theta_1-(1-\kappa)-\kappa \theta_3]D_{22}}{1+\kappa}.
\label{thermGKRO_5} 
\end{align}
This result implies that the time-reversal symmetry is broken for the thermal microswimmer 
when the temperatures are different.

\section{Effective odd elasticity of the thermal micromachine}
\label{Sec:Oddelasticity}

So far, we have discussed the statistical properties of the odd and thermal micromachines, 
and obtained their time-correlation functions to discuss the respective Green-Kubo relations. 
Although these two stochastic models are different, it is worthwhile to consider the effective odd 
elasticity of the thermal micromachine.  
As we mentioned earlier, odd elasticity gives a quantitative measure of the work that can be 
extracted from active systems~\cite{Scheibner20,Fruchart23}, and nonequilibrium cyclic dynamics 
of micromachines can be quantified by effective odd elasticity~\cite{Yasuda21catalytic,YasudaOM22}.
Such an approach is possible for stochastic micromachines by looking at the violation of the 
time-reversal symmetry of the cross-correlation functions~\cite{YasudaTCF22,KobayashiODD23}.

For this purpose, we compare the Green-Kubo relations in Eqs.~(\ref{GKRO_5}) and 
(\ref{thermGKRO_5}), and define an effective odd elasticity $\lambda_{\rm t}$ ratio of the thermal 
micromachine as 
\begin{align} 
\lambda_{\rm t}= \frac{D_{22}}{6D}[\theta_1-(1-\kappa)-\kappa \theta_3].
\label{ROTE}
\end{align}
If we further assume $D_{22}=D$ and $\kappa=1$, we obtain 
\begin{align} 
\lambda_{\rm t}= \frac{T_1- T_3}{6T_2}.
\label{ROTEsimple}
\end{align}
This result clearly demonstrates that the effective odd elasticity of the thermal micromachine 
is purely determined by the temperature difference between the edge spheres. 
The above relations further indicate that $\lambda_{\rm t}$ is inversely proportional to 
the middle sphere temperature $T_2$.

It should be mentioned, however, that the above argument is not the only way to obtain the 
effective odd elasticity of the thermal micromachine. 
For example, one can also compare the average velocities of the two micromachines as given in 
Eqs.~(\ref{AveV02}) and (\ref{TAveV1}) or the entropy production rates in Eqs.~(\ref{AEPRO1}) 
and (\ref{TAEPR1}).
Although the comparison of different quantities leads to a slightly modified numerical factor, the 
effective odd elasticity is always proportional to the temperature difference among the spheres, 
as shown in Eq.~(\ref{ROTEsimple}).  
This is because both the odd elasticity and the temperature difference lead to the violation 
of the time-reversal symmetry.
One of the advantages of using the Green-Kubo relations to define $\lambda_{\rm t}$ is that they 
contain information on the cross-correlation functions over the whole time scale.

\section{Summary and discussion}
\label{Sec:Dis}

In this paper, we have calculated the time-correlation functions of the two different stochastic three-sphere 
micromachines, i,e., the odd micromachine and the thermal micromachine. 
For the both models, the cross-correlation functions of the two springs contain the antisymmetric part, which is 
a direct consequence of the broken time-reversal symmetry of the micromachines. 
For the odd micromachine, the antisymmetric part of the correlation function is proportional to the odd elastic 
constant [see Eq.~(\ref{DTCFO_AB2})], whereas it is proportional to the temperature difference among 
the spheres [see Eq.~(\ref{TCFT_AB^A})].
We have also obtained the average entropy production rate and the Green-Kubo relations for the two 
micromachines.
A comparison of these results allows us to discuss the effective odd elasticity of the thermal micromachine 
[see Eq.~(\ref{ROTE})], even though the odd elasticity is not explicitly included in this model.    
For the thermal micromachine, the effective odd elasticity is proportional to the temperature difference 
between the spheres, which causes an internal heat flow and leads to directional locomotion.

Evaluation of effective odd elasticity was previously discussed for a model micromachine driven by 
catalytic chemical reactions~\cite{KobayashiODD23}. 
We calculated the time-correlation functions of the structural variables and analyzed them in 
terms of Langevin dynamics with effective odd elasticity~\cite{YasudaTCF22}. 
It was also shown that the odd elasticity is directly related to the quantity called 
\textit{nonreciprocality} of a micromachine. 
For a deterministic micromachine undergoing cyclic motions, the nonreciprocality $R$ is defined 
by~\cite{Yasuda21}  
\begin{align}
R = \frac{1}{2} \oint dt \, (u_{\rm A} \dot{u}_{\rm B} - \dot{u}_{\rm A} u_{\rm B}),
\label{nonreciprocality}
\end{align}
where the integral is taken over one cycle, and $R$ represents the area enclosed by the trajectory 
in the configuration space~\cite{Shapere89}. 
For stochastic micromachines, the average nonreciprocality $\langle R \rangle$ is essentially given by 
the time derivative of the antisymmetric part of the extension-extension correlation function, 
such as given by Eq.~(\ref{DTCFO_AB2}) that is proportional to the odd elasticity.
Hence, the effective odd elasticity introduced in Eq.~(\ref{ROTEsimple}) directly characterizes the nonreciprocality 
of the thermal micromachine.

Recently, Hargus \textit{et al.}\ discussed odd diffusivity in two-dimensional chiral active matter and 
derived the Green-Kubo relation for the odd diffusion coefficient~\cite{Hargus20prl}.
A similar relation was also obtained for the odd mobility of a passive tracer in a chiral active 
fluid~\cite{Poggioli23}.
In their results, both odd diffusivity and odd mobility result from the broken time-reversal symmetry
in active systems and are related to the antisymmetric part of the cross-correlation functions, which 
is similar to our result.
In our stochastic micromachines, the integrals of the auto-correlations vanish as in Eqs.~(\ref{GKRO_3}) 
and (\ref{GKRO_3therm}) because the spheres are connected by the springs. 
Although the motions of the micromachines are one-dimensional in our models, the configurational space 
spanned by $u_{\rm A}$ and $u_{\rm B}$ are two-dimensional. 
It is useful to consider effective odd elasticity in such a configurational space, and the obtained 
Green-Kubo relations for the cross-correlation functions in Eqs.~(\ref{GKRO_5}) and (\ref{thermGKRO_5})
are directly proportional to such effective odd elasticity.
We also note that the oscillatory trajectory seen in the chiral random walk~\cite{Hargus20prl} is 
analogous to the oscillatory behaviors of the time-correlation functions in this paper due to the
odd elasticity.

In this work, the elastic constant matrix for the odd micromachine was given by Eq.~(\ref{ECK}) in which 
we took into account the difference between $K_{\rm A}^{\rm e}$ and $K_{\rm B}^{\rm e}$. 
However, the most general form of the elastic constant matrix can take the form 
\begin{align} 
\mathbf{K}
=\left(                 
\begin{array}{cc}   
K_{\rm A}^{\rm e} & K^{\rm e}_{\rm AB}+ K^{\rm o} \vspace{1ex} 
\\ 
K^{\rm e}_{\rm AB} -K^{\rm o} & K_{\rm B}^{\rm e} \\
\end{array}
\right),
\label{ECKgen}     
\end{align}
where we have introduced another even elastic constant $K^{\rm e}_{\rm AB}$ in the off-diagonal 
components~\cite{KobayashiODD23}.
The elastic constant $K^{\rm e}_{\rm AB}$ can induce amplitude asymmetry between the two springs~\cite{Lin23}. 
Although the time-correlation functions can be easily calculated in the presence of $K^{\rm e}_{\rm AB}$,
Eq.~(\ref{ECK}) is sufficient to compare with the thermal micromachine.

To calculate the time-correlation functions, we have neglected hydrodynamic interactions acting between 
the spheres. 
This is justified because hydrodynamic interactions are higher-order contributions for the correlation functions.
If hydrodynamic interactions are taken into account, the results in this work will be accompanied by correction 
terms that are proportional to $a/\ell$ within the lowest-order expansion~\cite{Sou19,Sou21}.
It should be mentioned again, however, that hydrodynamic interactions are necessary for the locomotion of 
the three-sphere micromachines.

\begin{acknowledgments}

We thank R. Garc\'ia-Mill\'an, L.-S.\ Lin, and Z.\ Xiong for useful discussion.
Z.H.\, L.H.\, and S.K.\ acknowledge the support by the National Natural Science Foundation of China 
(Nos.\ 12104453, 22273067, 12274098, and 12250710127).
K.Y.\ acknowledges the support by a Grant-in-Aid for JSPS Fellows (Grants No.\ 22KJ1640) from the JSPS.
S.K.\ acknowledges the startup grant of Wenzhou Institute, 
University of Chinese Academy of Sciences (No.\ WIUCASQD2021041). 
K.Y\ and S.K.\ acknowledge the support by the Japan Society for the Promotion of Science (JSPS) Core-to-Core 
Program ``Advanced core-to-core network for the physics of self-organizing active matter" (No.\ JPJSCCA20230002).
J.L.\ and Z.Z.\ contributed equally to this work.
\end{acknowledgments}

\appendix
\begin{widetext}
\section{Extension-extension correlation functions of the odd three-sphere micromachine}
\label{App:A}

In this Appendix, we show the derivation of the extension-extension correlation functions in 
Eqs.~(\ref{DTCFO_AA})-(\ref{DTCFO_AB2}).
Using Eqs.~(\ref{FreqU_A}) and (\ref{FreqU_B}), we first obtain the three correlation functions
$\langle u_{\rm A}(\omega)u_{\rm A}(\omega')\rangle$, $\langle u_{\rm B}(\omega)u_{\rm B}(\omega')\rangle$,  
and $\langle u_{\rm A}(\omega)u_{\rm B}(\omega')\rangle$ in the frequency domain.
By performing their inverse Fourier transform, we obtain 
\begin{align}
\langle u_{\rm A}(t)u_{\rm A}(0)\rangle
&=\int_{-\infty}^{\infty}\dfrac{d\omega}{2\pi} 
\int_{-\infty}^{\infty}\dfrac{d\omega'}{2\pi} 
\left\langle u_{\rm A}(\omega)u_{\rm A}(\omega')\right\rangle 
e^{{\rm i}\omega t}\nonumber\\
&=\dfrac{D}{\pi\tau^2}\int_{-\infty}^{+\infty}d\omega \,
\dfrac{6(\kappa^2-\kappa\lambda+\lambda^2)+2\omega^2\tau^2}
{\left[\omega^2-2{\rm i}(1+\kappa)\omega/\tau-3(\kappa+\lambda^2)/\tau^2\right]
\left[\omega^2+2{\rm i}(1+\kappa)\omega/\tau-3(\kappa+\lambda^2)/\tau^2\right]}
e^{{\rm i}\omega t},\label{OTCFI_1}\\ 
\langle u_{\rm B}(t)u_{\rm B}(0)\rangle
&=\int_{-\infty}^{\infty}\dfrac{d\omega}{2\pi}
\int_{-\infty}^{\infty}\dfrac{d\omega'}{2\pi}
\left\langle u_{\rm B}(\omega)u_{\rm B}(\omega')\right\rangle 
e^{{\rm i}\omega t}\nonumber\\
&=\dfrac{D}{\pi\tau^2}\int_{-\infty}^{+\infty}d\omega \,
\dfrac{6(1+\lambda+\lambda^2)+2\omega^2\tau^2}
{\left[\omega^2-2{\rm i}(1+\kappa)\omega/\tau-3(\kappa+\lambda^2)/\tau^2\right]
\left[\omega^2+2{\rm i}(1+\kappa)\omega/\tau-3(\kappa+\lambda^2)/\tau^2\right]}
e^{{\rm i}\omega t},\label{OTCFI_2}\\ 
\langle u_{\rm A}(t)u_{\rm B}(0)\rangle
&=\int_{-\infty}^{\infty}\dfrac{d\omega}{2\pi}
\int_{-\infty}^{\infty}\dfrac{d\omega'}{2\pi}
\left\langle u_{\rm A}(\omega)u_{\rm B}(\omega')\right\rangle 
e^{{\rm i}\omega t}\nonumber\\
&=\dfrac{D}{\pi\tau^2}\int_{-\infty}^{+\infty}d\omega \,
\dfrac{3[\kappa-2(1-\kappa)\lambda-\lambda^2]+6\rm{i}\lambda\omega\tau-\omega^2\tau^2}
{\left[\omega^2-2{\rm i}(1+\kappa)\omega/\tau-3(\kappa+\lambda^2)/\tau^2\right]
\left[\omega^2+2{\rm i}(1+\kappa)\omega/\tau-3(\kappa+\lambda^2)/\tau^2\right]}
e^{{\rm i}\omega t},\label{OTCFI_3}
\end{align}
where $D=k_{\rm B}T/(6\pi\eta a)$.

The above integrand functions have four poles 
$\omega_1=(\mu + {\rm i}\nu)/\tau$, 
$\omega_2=(-\mu + {\rm i}\nu)/\tau$, 
$\omega_3=-(\mu + {\rm i}\nu)/\tau$, and
$\omega_4=(\mu - {\rm i}\nu)/\tau$, where
$\mu=\sqrt{3\lambda^2-(1-\kappa+\kappa^2)}$ and $\nu=1+\kappa$.
When $3\lambda^2-(1-\kappa+\kappa^2)>0$, as shown in Fig.~\ref{Fig:Residue}(a), 
the integrand functions have four first-oder poles in the complex $\omega$-plane.
When $3\lambda^2-(1-\kappa+\kappa^2)=0$, as shown in Fig.~\ref{Fig:Residue}(b), 
$\omega_1$ and $\omega_3$  become second-oder poles because 
$\omega_1=\omega_2=\rm{i}\nu/\tau$ and $\omega_3=\omega_4=-\rm{i}\nu/\tau$.
When $3\lambda^2-(1-\kappa+\kappa^2)<0$, as shown in Fig.~\ref{Fig:Residue}(c),
the integrand functions have four first-oder poles on the imaginary axis. 
Using the residue theorem, we can perform the above integrals to obtain 
Eqs.~(\ref{DTCFO_AA})-(\ref{DTCFO_AB2}).
As indicated in Fig.~\ref{Fig:Residue}, the integration path should be counter-clockwise
and clockwise for $t>0$ and $t<0$, repectively.

\begin{figure*}[tb]
\centering
\includegraphics[scale=0.8]{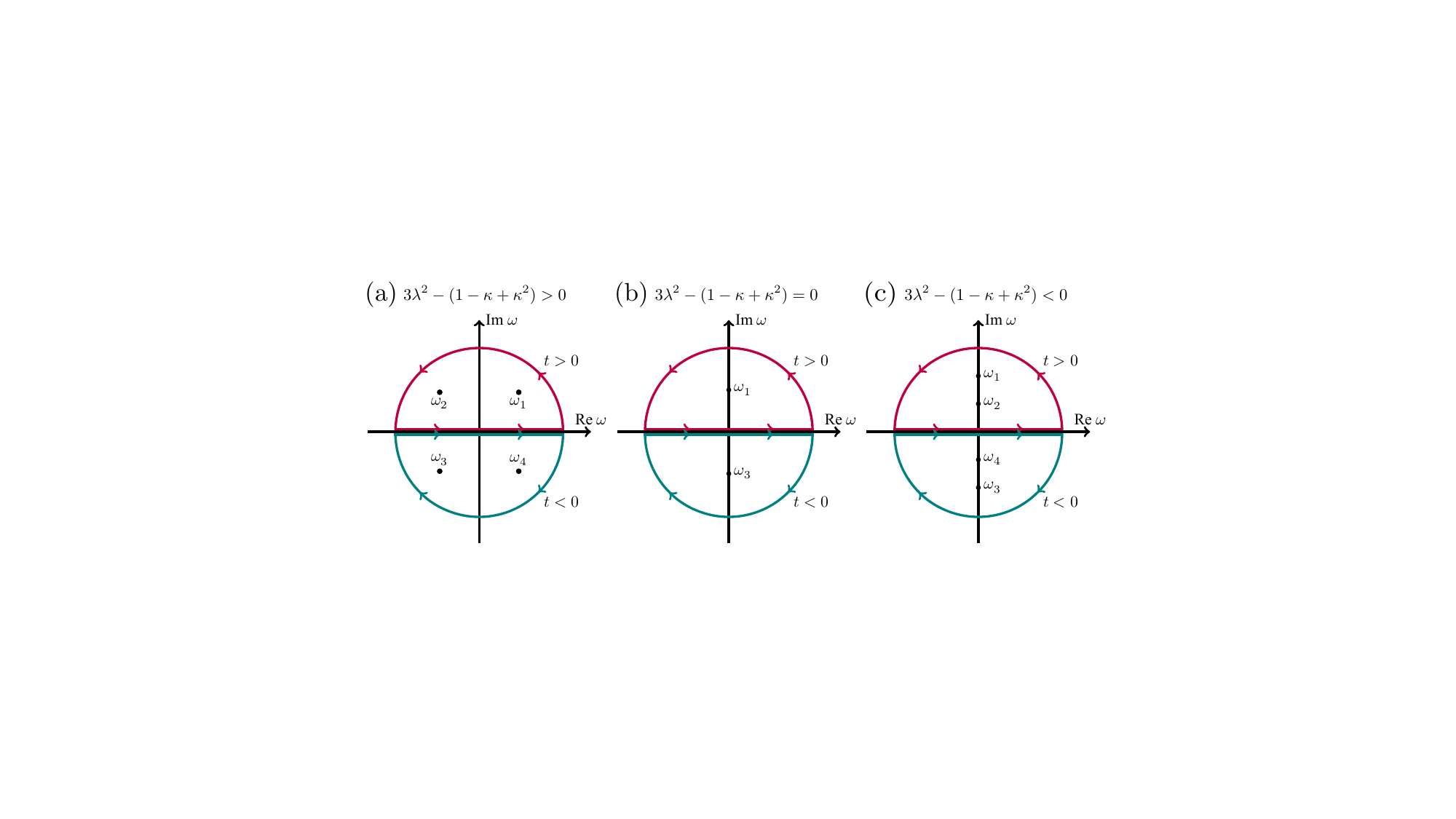}
\caption{The poles and the integration paths on the complex $\omega$-plane.}
\label{Fig:Residue}
\end{figure*}

\section{Extension-extension correlation functions of the thermal three-sphere micromachine}
\label{App:B}

In this Appendix, we show the derivation of the extension-extension correlation functions in 
Eqs.~(\ref{TCFT_AA})-(\ref{TCFT_AB^A}).
They are given by 
\begin{align}
{\langle u_{\rm A}(t)u_{\rm A}(0)\rangle}_{\rm t}
&=\dfrac{D_{22}}{\pi\tau^2}\int_{-\infty}^{+\infty}d\omega \,
\dfrac{(4\theta_1+1+\theta_3)\kappa^2+(1+\theta_1)\omega^2\tau^2}
{\left[\omega^2-2{\rm i}(1+\kappa)\omega/\tau
-3\kappa/\tau^2\right]\left[\omega^2
+2{\rm i}(1+\kappa)\omega/\tau
-3\kappa/\tau^2\right]}e^{{\rm i}\omega t},
\label{GTCFOTM_AA}\\
{\langle u_{\rm B}(t)u_{\rm B}(0)\rangle}_{\rm t}
&=\dfrac{D_{22}}{\pi\tau^2}\int_{-\infty}^{+\infty}d\omega \,
\dfrac{(\theta_1+1+4\theta_3)+(1+\theta_3)\omega^2\tau^2}
{\left[\omega^2-2{\rm i}(1+\kappa)\omega/\tau
-3\kappa/\tau^2\right]\left[\omega^2
+2{\rm i}(1+\kappa)\omega/\tau
-3\kappa/\tau^2\right]}e^{{\rm i}\omega t},
\label{GTCFOTM_BB}\\
{\langle u_{\rm A}(t)u_{\rm B}(0)\rangle}_{\rm t}
&=\dfrac{D_{22}}{\pi\tau^2}\int_{-\infty}^{+\infty}d\omega \, 
\dfrac{(2\theta_1-1+2\theta_3)\kappa+{\rm i}[\theta_1-(1-\kappa)
-\kappa\theta_3]\omega\tau-\omega^2\tau^2}
{\left[\omega^2-2{\rm i}(1+\kappa)\omega/\tau
-3\kappa/\tau^2\right]\left[\omega^2
+2{\rm i}(1+\kappa)\omega/\tau
-3\kappa/\tau^2\right]}e^{{\rm i}\omega t},
\label{GTCFOTM_AB}
\end{align}
where $D_{22}=k_{\rm B}T_2/(6\pi\eta a)$.
Following the similar procedure as in Appendix~\ref{App:A}, we obtain Eqs.~(\ref{TCFT_AA})-(\ref{TCFT_AB^A}).
Due to the absence of the odd elasticity, however, the integrand functions always have four first-oder poles on the 
imaginary axis, as in Fig.~\ref{Fig:Residue}(c).

\section{Velocity-velocity correlation functions of the odd three-sphere micromachine}
\label{App:C}

As mentioned in the text, the velocity-velocity correlation functions can be calculated by using Eq.~(\ref{RLEq}) as 
\begin{align}
\langle \dot{u}_{\alpha}(t)\dot{u}_{\beta}(0)\rangle
&=\langle[\Gamma_{\alpha\gamma}u_{\gamma}(t)+\Xi_{\alpha}(t)]
[\Gamma_{\beta\delta}u_{\delta}(0)+\Xi_{\beta}(0)] \rangle.
\label{GVTCF}
\end{align}
The correlations between the displacement and noise, $\langle u_{\alpha}(t) \Xi_{\beta}(0) \rangle$, can also 
be calculated straightforwardly.

For the odd micromachine, the expressions of the velocity-velocity correlation functions defined 
in Eq.~(\ref{RVDTCFO}) are given as follows:
\begin{align}
\psi_{\rm{AA}}^{\rm s}(t)
&=4\tau\delta(t)+\frac{6\lambda^2-3\kappa\lambda-(4+5\kappa+\kappa^2)}{1+\kappa}
\cos\left(\mu t/\tau \right)e^{-(1+\kappa)\vert t \vert/\tau}
-\frac{12\lambda^2-3\kappa\lambda-(4-\kappa+\kappa^2)}{\mu}
\sin\left(\mu \vert t \vert/\tau\right) e^{-(1+\kappa) \vert t \vert /\tau},
\label{SVTCFO_AA}\\ 
\psi_{\rm{BB}}^{\rm s}(t)
&=4\tau\delta(t)+\frac{6\lambda^2+3\lambda-(1+5\kappa+4\kappa^2)}{1+\kappa}
\cos\left(\mu t/\tau\right)e^{-(1+\kappa)\vert t \vert/\tau}
-\frac{12\lambda^2+3\lambda-(1-\kappa+4\kappa^2)}{\mu}
\sin\left(\mu \vert t \vert/\tau\right)e^{-(1+\kappa)\vert t \vert /\tau},
\label{SVTCFO_BB} \\
\psi_{\rm{AB}}^{\rm s}(t) 
&= -2\tau\delta(t)
\nonumber \\
& -\frac{3\lambda^2+3(1-\kappa)\lambda-2(1+2\kappa+\kappa^2)}{1+\kappa}
\cos\left(\mu t/\tau\right)e^{-(1+\kappa) \vert t \vert/\tau}
+\frac{6\lambda^2+3(1-\kappa)\lambda-(2+\kappa+2\kappa^2)}{\mu}
\sin\left(\mu \vert t \vert/\tau\right)e^{-(1+\kappa) \vert t \vert/\tau},
\label{VTCFO_AB1} \\
\psi_{\rm{AB}}^{\rm a}(t) 
& =\begin{cases}
-6\lambda \cos\left(\mu t/\tau\right)e^{-(1+\kappa)t/\tau} 
-\dfrac{3\lambda[3\lambda^2-(2+\kappa+2\kappa^2)]}{(1+\kappa)\mu}
\sin\left( \mu t/\tau\right)e^{-(1+\kappa)t/\tau} & (t>0), \vspace{1ex}
\\
6\lambda \cos\left(\mu t/\tau\right)e^{(1+\kappa)t/\tau}
-\dfrac{3\lambda[3\lambda^2-(2+\kappa+2\kappa^2)]}{(1+\kappa)\mu}
\sin\left( \mu t/\tau\right)e^{(1+\kappa)t/\tau} & (t<0),
\label{VTCFO_AB2} 
\end{cases}
\end{align}
where $\mu=\sqrt{3\lambda^2-(1-\kappa+\kappa^2)}$.

\section{Velocity-velocity correlation functions of the thermal three-sphere micromachine}
\label{App:D}

\begin{figure*}[tb]
\centering
\includegraphics[scale=0.58]{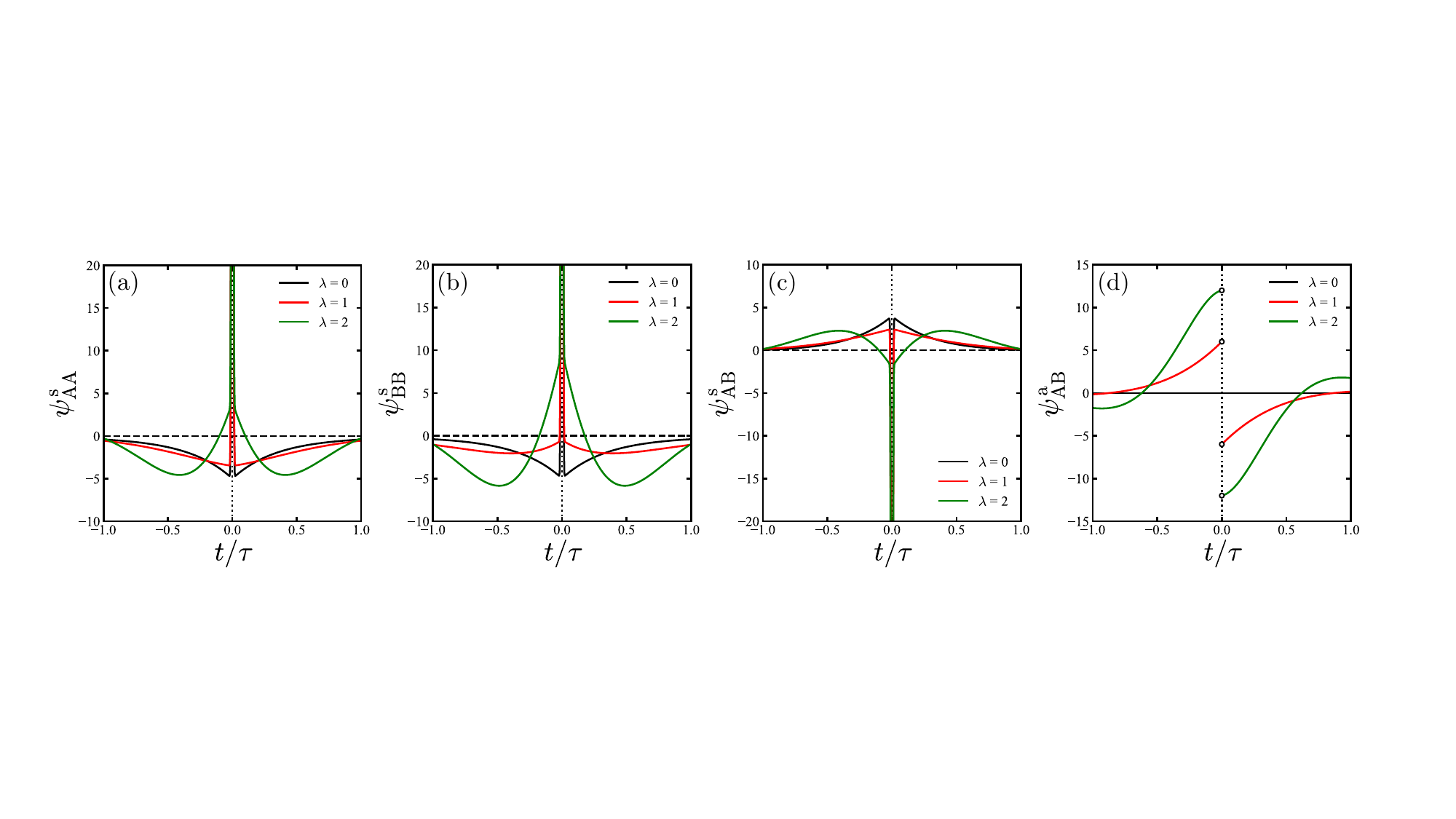}
\caption{Plots of the scaled velocity-velocity correlation functions 
(a) $\psi_{\rm{AA}}^{\rm s}$, (b) $\psi_{\rm{BB}}^{\rm s}$, (c) $\psi_{\rm{AB}}^{\rm s}$, and 
(d) $\psi_{\rm{AB}}^{\rm a}$ as a function of $t/\tau$ for $\kappa=1$.
The dimensionless odd elasticity is chosen as $\lambda=0$ (black), $1$ (red), and $2$ (green).
}
\label{Fig:DTCFOimodd}
\end{figure*}

\begin{figure*}[tb]
\centering
\includegraphics[scale=0.58]{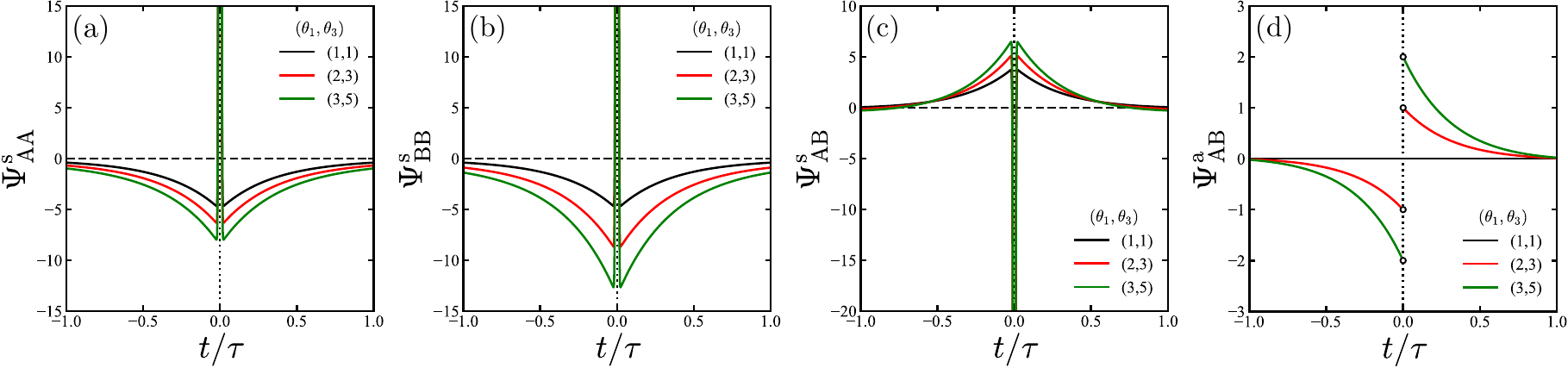}
\caption{Plots of the scaled velocity-velocity correlation functions 
(a) $\Psi_{\rm{AA}}^{\rm s}$, (b) $\Psi_{\rm{BB}}^{\rm s}$, (c) $\Psi_{\rm{AB}}^{\rm s}$, and 
(d) $\Psi_{\rm{AB}}^{\rm a}$ as a function of $t/\tau$ for $\kappa=1$.
The dimensionless temperature ratios between the spheres are chosen as 
$(\theta_1,\theta_3)=(1,1)$ (black), 
$(2,3)$ (red), and 
$(3,5)$ (green).}
\label{Fig:DTCFTthermal}
\end{figure*}

For the thermal micromachine, the expressions of the velocity-velocity correlation functions defined 
in Eq.~(\ref{thermRVDTCFO}) are given as follows:
\begin{align}
\Psi_{\rm{AA}}^{\rm s}(t) 
& = 2(\theta_1+1)\tau\delta(t)
\nonumber \\
& -\frac{(4+5\kappa)\theta_1+(4+5\kappa+3\kappa^2)-\kappa^2\theta_3}{2(1+\kappa)}
\cosh\left(\rho t/\tau \right )e^{-(1+\kappa) \vert t \vert/\tau}
- \frac{(\kappa-4)\theta_1-(4-\kappa+3\kappa^2)+\kappa^2\theta_3}{2\rho}
\sinh \left(\rho \vert t \vert/\tau \right)
e^{-(1+\kappa) \vert t \vert/\tau},
\label{PsiAA}
\\
\Psi_{\rm{BB}}^{\rm s}(t) 
& = 2(1+\theta_3)\tau\delta(t)
\nonumber \\
& -\frac{-\theta_1+(3+5\kappa+4\kappa^2)+\kappa(5+4\kappa)\theta_3}{2(1+\kappa)}
\cosh\left(\rho t/\tau \right)e^{-(1+\kappa) \vert t \vert/\tau}
- \frac{\theta_1-(3-\kappa+4\kappa^2)+\kappa(1-4\kappa)\theta_3}{2\rho}
\sinh\left(\rho \vert t \vert/\tau \right)
e^{-(1+\kappa) \vert t \vert/\tau}, 
\label{PsiBB}
\\
\Psi^{\rm s}_{\rm{AB}}(t) 
&= -2\tau\delta(t)
+\frac{\kappa \theta_1+2(1+\kappa+\kappa^2)+\kappa \theta_3}{1+\kappa}
\cosh\left(\rho t/\tau \right)e^{-(1+\kappa) \vert t \vert/\tau}
-\frac{\kappa \theta_1+(2-\kappa+2\kappa^2)+\kappa \theta_3}{\rho}
\sinh\left(\rho \vert t \vert/\tau \right)
e^{-(1+\kappa) \vert t \vert/\tau},
\label{PsiABeven}
\\
\Psi^{\rm a}_{\rm{AB}}(t) 
& =\begin{cases}
-[\theta_1-(1-\kappa)-\kappa \theta_3]
\cosh\left(\rho t/\tau\right)e^{-(1+\kappa)t/\tau} 
+\dfrac{(2+\kappa+2\kappa^2)[\theta_1-(1-\kappa)-\kappa \theta_3]}{2(1+\kappa)\rho}
\sinh\left(\rho t/\tau\right)e^{-(1+\kappa)t/\tau} & (t>0), \vspace{1ex}
\\
[\theta_1-(1-\kappa)-\kappa \theta_3]
\cosh\left(\rho t/\tau\right)e^{(1+\kappa)t/\tau} 
+\dfrac{(2+\kappa+2\kappa^2)[\theta_1-(1-\kappa)-\kappa \theta_3]}{2(1+\kappa)\rho}
\sinh\left(\rho t/\tau\right)e^{(1+\kappa)t/\tau} & (t<0),
\end{cases}
\label{PsiABodd}
\end{align}
where $\rho=\sqrt{1-\kappa+\kappa^2}$.

\end{widetext}


\end{document}